\documentclass{webofc}
\usepackage[varg]{txfonts}   
\begin{document}
\title{Collectivity in small systems - Initial state vs. final state effects}
%
%

\author{
\firstname{Moritz} \lastname{Greif} \inst{1}
\and
\firstname{Carsten} \lastname{Greiner} \inst{1}
\and
\firstname{Bj\"orn} \lastname{Schenke}\inst{2}
\and
\firstname{S\"{o}ren} \lastname{Schlichting}\inst{3}\fnsep\thanks{\email{sslng@uw.edu}}
\and
\firstname{Zhe} \lastname{Xu}\inst{4}
}

\institute{
Institut f\"ur Theoretische Physik, Johann Wolfgang Goethe-Universit\"at, Max-von-Laue-Str.\ 1, D-60438 Frankfurt am Main, Germany
\and 
Physics Department, Brookhaven National Laboratory, Upton, NY 11973, USA
\and
Department of Physics, University of Washington, Seattle, WA 98195-1560, USA 
\and
Department of Physics, Tsinghua University and Collaborative Innovation Center of Quantum Matter, Beijing 100084, China
}

\abstract{Observations of long rang azimuthal correlations in small collision systems (p+p/A) have triggered an enormous excitement in the heavy-ion community. However, it is presently unclear to what extent the experimentally observed correlations should be attributed to initial state momentum correlations and/or the final state response to the initial state geometry. We discuss how a consistent theoretical description of the non-equilibrium dynamics is important to address both effects within a unified framework and present first results from weakly coupled non-equilibrium simulations in \cite{Greif:2017bnr} to quantify the relative importance of initial state and final state effects based on theoretical calculations.}
\maketitle
\section{Introduction}
\label{intro}
Experimental measurements of long. range (in rel. rapidity $\Delta \eta$) azimuthal correlations (in rel. angle $\Delta \phi$) in high-energy proton-proton (p+p) and proton-nucleus (p+A) collisions, have revealed many interesting features that call for a deeper theoretical understanding~\cite{Dusling:2015gta}. Most strikingly, many features observed in these ``small system''  are qualitatively similar to previous observations in nucleus-nucleus (A+A) collisions, including e.g. the transverse momentum and hadron-species dependence of these correlations \cite{Abelev:2013wsa,Khachatryan:2014jra} as well as the fact that the observed correlations are ``collective'' in the sense that many particles are correlated with each other \cite{Chatrchyan:2013nka,Khachatryan:2015waa,Aad:2013fja}. However, there are also some important differences between the experimental results in proton-proton/nucleus and nucleus-nucleus collisions, including e.g. the fact that (so far) no evidence of jet-quenching phenomena has been reported in small systems.\\

Based on the apparent similarities and differences between small (p+p/A) and large (A+A) systems, different theoretical explanations have been explored which attribute the observed correlations either to initial state momentum correlations (see e.g. \cite{Dusling:2012cg,Dusling:2013qoz,Dumitru:2014yza,Dusling:2015rja,Schenke:2016lrs,Dusling:2017aot}) or the final state response to the event geometry (see e.g. \cite{Bozek:2011if,Qin:2013bha,Werner:2013ipa,Schenke:2014zha,Mantysaari:2017cni,Weller:2017tsr}). Since the format of this proceeding is not suitable to provide an exhaustive review of the current state of the art of these calculations,\footnote{We refer the interested reader to~\cite{Dusling:2015gta,Schlichting:2016sqo} for detailed reviews on these subjects.} we will focus on the key challenge of quantifying the relative importance of these two effects. Based on a short outline of the general expectations in Sec.~\ref{sec-CARTOON}, we present first results addressing this questions within a weak coupling approach to the underlying non-equilibrium dynamics in Sec.~\ref{sec-IPGLAMPS}, followed by some concluding remarks and perspectives for future studies in Sec.~\ref{sec-conc}.    

\begin{figure}[t!]
\centering
\includegraphics[width=\textwidth,clip]{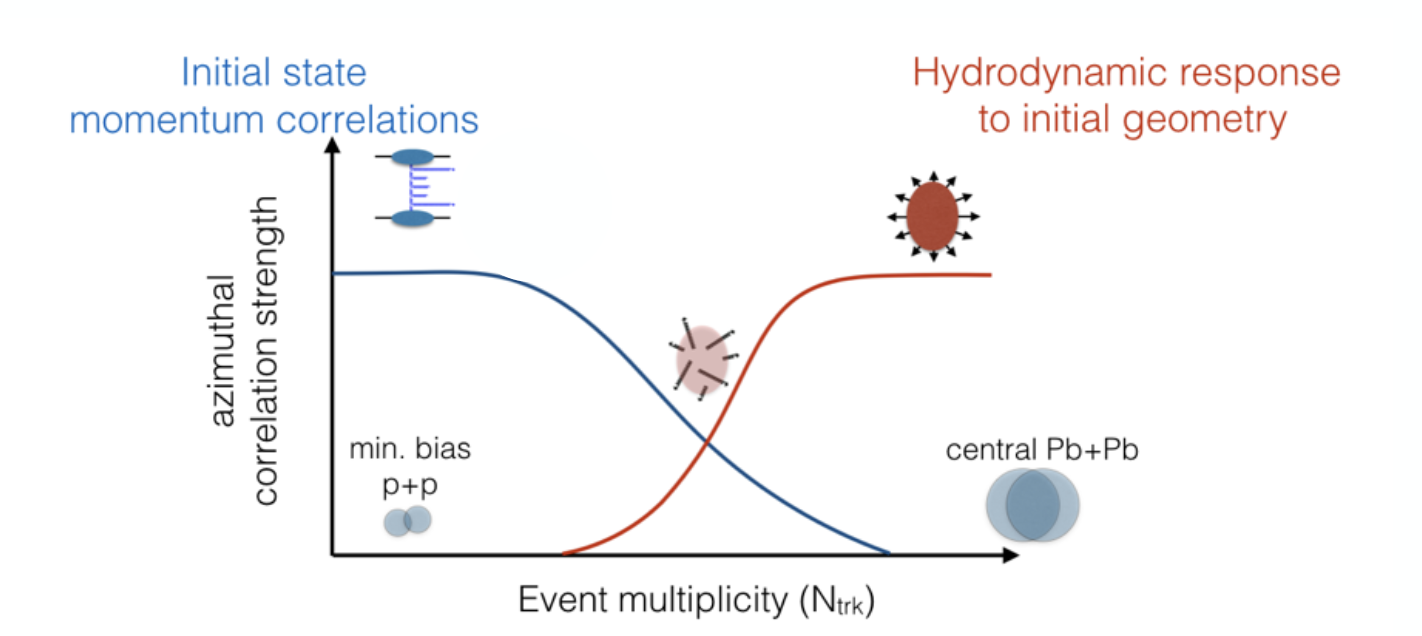}
\caption{Illustration of the expected relative strength of different sources of long range azimuthal correlations: (blue) initial state momentum correlations (red) final state response to initial state geometry. Beyond the asymptotic limits ($N_{trk}\to0$ and $N_{trk}\to \infty$), one expects long range azimuthal correlations  to be sensitive to the non-equilibrium dynamics of the medium. Fig. from \cite{Schlichting:2016xmj,Schlichting:2016sqo}}
\label{fig-1}       
\end{figure}

\section{Quantifying the importance initial state vs. final state effects}
\label{sec-main}
\subsection{General expectations}
\label{sec-CARTOON}
Generally speaking, it is important to take into account both initial state and final state effects in order to obtain a robust theoretical description across a wide kinematic range all the way from low multiplicity p+p ($dN_{p+p}^{min. bias}/dy \sim 5$)  to high-multiplicity A+A collisions ($dN_{Pb+Pb}^{0-5\%}/dy \sim 1500$). While one naturally expects initial state effects to dominate correlations in low mult. p+p collisions, where the opacity of the medium is small and final state interactions are subleading, the opacity of the medium in high mult. A+A collisions is very large such that even highly energetic jets loose a substantial amount of energy and azimuthal correlations between low momentum particles emerge pre-dominantly due to the hydrodynamic response of the medium to the initial state geometry. However, as illustrated in Fig.~\ref{fig-1}, this leaves a large range of possibilities where for the typical degrees of freedom the opacity of the medium is neither particularly small to be ignored nor particularly large for the medium to be described hydrodynamically. Hence one expects experimental observables to become sensitive to the non-equilibrium dynamics of the medium, and new theoretical developments are needed to describe long. range azimuthal correlations in this regime.  \\

\subsection{Non-equilibrium description of initial state \& final state effects}
\label{sec-IPGLAMPS}
Based on a weak-coupling picture, a consistent theoretical description of the non-equilibrium dynamics in high-energy collisions~\cite{Baier:2000sb,Berges:2013fga,Kurkela:2015qoa} can be achieved by matching a classical-statistical lattice description of particle production and early time dynamics (see e.g. ~\cite{Berges:2013fga}) to an effective kinetic description of the subsequent non-equilibrium dynamics (see e.g.~\cite{Kurkela:2015qoa}). Event-by-event simulations of the non-equilibrium dynamics of high-energy p+p,p+A and A+A collisions, can then be performed within this framework by
\begin{itemize}
\item[1)] ~Simulating particle production and early time dynamics ($\tau<0.2 fm/c$) based on the IP-Glasma model~\cite{Schenke:2012wb,Schenke:2012hg}
\item[2)] ~Extracting the phase space distribution of gluons $dN_g/d\eta d^2{\bf x} dy d^2{\bf p}$, which includes all relevant information about initial state momentum correlations\footnote{Event-by-event the initial state momentum correlations due to Bose enhancement of small-x gluons (Glasma graphs) factorize into products of (azimuthally anisotropic) single particle distribution, i.e. $d^2N_{2g}/d^2{\bf p}_1d^2{\bf p}_2 = dN_{g}/d^2{\bf p}_1 \times dN_{g}/d^2{\bf p}_2$ \cite{Dusling:2009ni,Lappi:2009xa,Schenke:2015aqa} such that the associated one body density $dN_g/d\eta d^2{\bf x} dy d^2{\bf p}$ contains all relevant information about such correlations.}  as well the initial state geometry 
\item[3)] ~Simulating final state re-scattering dynamics in a parton cascade (BAMPS)~\cite{Xu:2004mz,Xu:2007aa,Xu:2014ega} based on pQCD matrix elements for $2<->2$ and $2<->3$ interactions \cite{Fochler:2013epa,Uphoff:2014cba}
\end{itemize}
and  details of the implementation are discussed in~\cite{Greif:2017bnr}. By investigating the time evolution of azimuthal correlations of partons over the course of the non-equilibrium evolution, one can then start to assess the relative importance of initial state and final state effects from a theoretical perspective. \\

\begin{figure}[t!]
\centering
\includegraphics[width=\textwidth,clip]{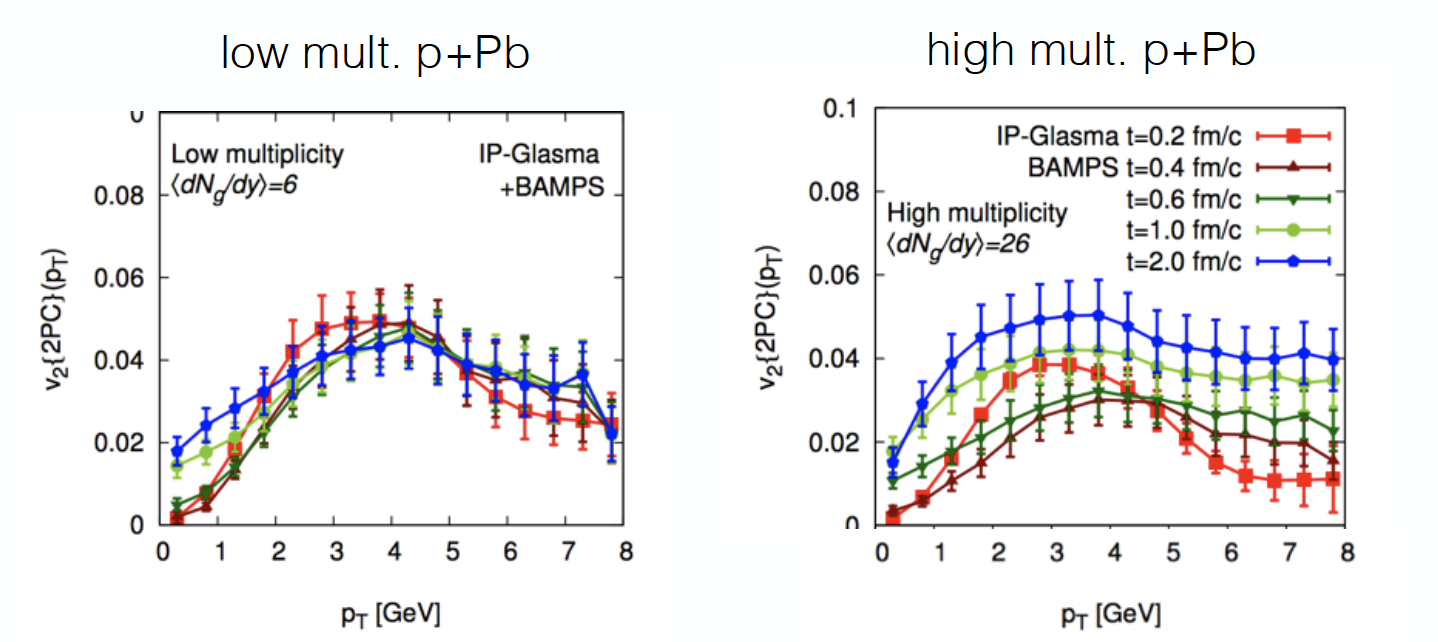}
\caption{Comparison of the time evolution of the (two-particle) azimuthal correlation coefficient $v_2^{gluon}$ in low. multiplicty (left) and high multiplicity (right) p+Pb collisions. Fig. from \cite{Greif:2017bnr}}
\label{fig-2}       
\end{figure}

Some results from~\cite{Greif:2017bnr} for azimuthal correlations in p+Pb collisions are summarized in Fig.~\ref{fig-2}, where the time evolution of the gluon $v_{2}(p_T)$ in low multiplicity (left) and high multiplicity (right) events is compared. Despite the fact that sizable initial state momentum correlations are present at $t=0.2 fm/c$ in both high and low multipiclity events (also see \cite{Lappi:2009xa,Schenke:2015aqa}), one observes a strong modification of the correlations in high multiplicity events ($\left( dN_g/dy\right)/ \langle dN_g/dy \rangle  > 2.5$), where most of the initial state correlations get destroyed on a time scale $\sim 1 fm/c$ while new final state induced correlations are build up simultaneously in response to the system geometry.\footnote{A careful analysis reveals that the final state induced correlations are strongly correlated with the event geometry~\cite{Greif:2017bnr}.} Conversely, in low multiplicity events  ($0.5 < \left( dN_g/dy\right)/ \langle dN_g/dy \rangle  < 1$)  final state re-scattering only affects correlations at very low momenta $p_T \lesssim 2 {\rm GeV}$; while initial state momentum correlations at higher momenta $p_T \gtrsim 2 {\rm GeV}$ persist throughout the evolution. Based on the discussion in Sec.~\ref{sec-CARTOON} it is natural that the differences between the two event classes can be attributed to a larger number of large angle scatterings in high multiplicity events. However, it is at least somewhat surprising that even in high multiplicity events an average number of $\approx 1$ large angle scattering is sufficient to build up a geometric response, although similar observations have been reported previously in \cite{He:2015hfa}.

\begin{figure}[t!]
\centering
\sidecaption
\includegraphics[width=8cm,clip]{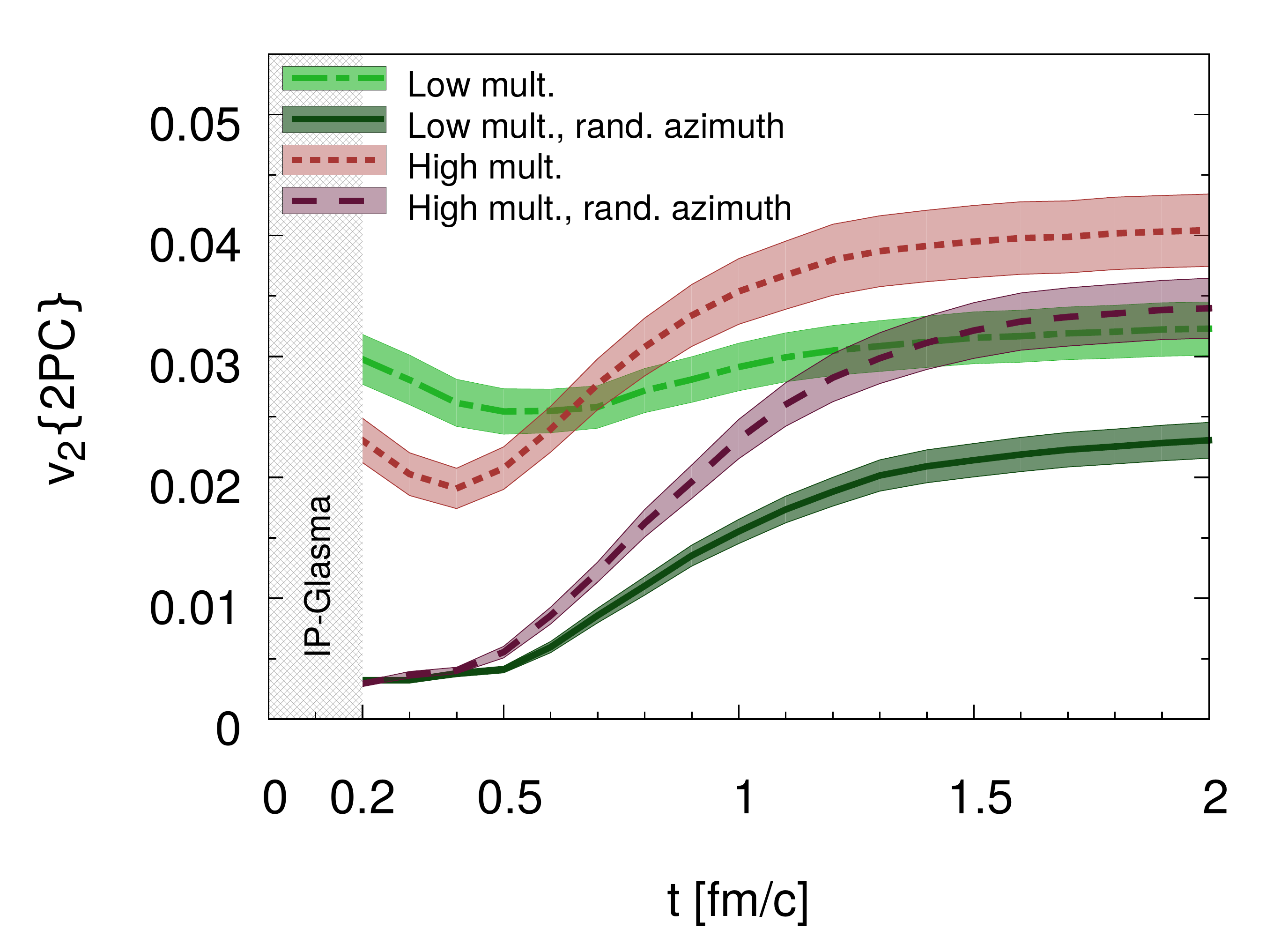}
\caption{Evolution of the $p_T$ integrated (gluon) $v_2$ for low (green) and high (red) multiplicity p+Pb events. Effect of initial state correlations on $v_2$ can be inferred from comparison of full simulations with ``rand. azimuth'' curves, where initial state momentum correlations have been artificially removed. Fig. from \cite{Greif:2017bnr}}
\label{fig-3}       
\end{figure}

Even though the relative importance of initial state and final state effects depends strongly on the momenta under consideration, an approximate measure of the different effects can be obtained based on the $p_T$ integrated $v_{2}$ which is depicted in Fig.~\ref{fig-3}. One finds that the time evolution of $v_2$ can be characterized by three distinct stages, where for $t \lesssim 0.2 fm/c$ correlations are dominated by initial state, before for $0.2 fm/c \lesssim t \lesssim 0.5 fm/c$ scatterings partially destroy initial state correlations until for $0.5 \lesssim t \lesssim 1.0 fm/c$ new correlations build up in response to event geometry. By comparing the results of full event-by-event simulations, with those where initial state correlations have been artificially removed by randomizing the azimuthal angles of the initial state  gluons (``rand. azimuth''), one can attempt to further disentangle initial state and final state contributions. Even though geometric response ultimately dominates $v_2$ at low $p_T$, the results in Fig.~\ref{fig-3} indicate that there are still sizable effects of initial state correlations (even on the $p_T$ integrated $v_2$) on the oder of $\sim25\%$ for high multiplicity and $\sim50\%$ for low multiplicity events.\\

\section{Conclusions \& Outlook}
\label{sec-conc}
Experimental observations of pronounced azimuthal correlations in small systems have started to challenge our current understanding of the space time evolution of high-energy collisions. So far two lines of theoretical explanations have been developed, which attribute the observed correlations either to intrinsic momentum correlations in the initial state or a collective expansion of the medium driven by final state interactions, while typically neglecting the other effect. Despite a variety of successful phenomenological applications in both frameworks, both of these approaches fall short in consistently describing the underlying dynamics over a wide range of multiplicities, where one expects a gradual change from an initial state dominated to a final state dominated scenario. Hence it is important to develop new theoretical tools, which can properly address the features of the non-equilibrium dynamics interpolating between these two extremes. So far, first calculations in this spirit \cite{Greif:2017bnr} indicate that both initial state and final state effects can be quantitatively important for two-particle correlations in p+Pb collisions. Based on such a consistent theoretical description of the non-equilibrium evolution, it would also be interesting to explore in more detail to what extent specific observables, such as e.g. the existence of (flow-like) higher order cumulants of $v_2$ at low momenta or the survival of (jet-like) back-to-back correlations a higher momenta, are indicative of the dominance of one of the underlying physics mechanisms. However, these studies are still in their infancies and further theoretical progress will be required to clarify these points.\\

\textit{Acknowledgements:} We would like to thank Kevin Dusling, Tuomas Lappi, Aleksas Mazeliauskas, Jean-Francois Paquet, Derek Teaney, Prithwish Tribedy and Raju Venugopalan for insightful discussions and fruitful collaborations related to this work. S.S. and B.S. acknowledge support by the U.S. Department of Energy, Office of Science, Office of Nuclear Physics under Grant DE-FG02-97ER41014 (S.S.) and DE-SC0012704 (B.S.). M.G., and C.G. acknowledge support by the Deutsche Forschungsgemeinschaft (DFG) through the grant CRC-TR 211 "Strong-interaction matter under extreme conditions". Z.X. was supported by the National Natural Science Foundation of China under Grants No. 11575092 and No. 11335005, and the Major State Basic Research Development Program in China under Grants No. 2014CB845400 and No. 2015CB856903.  Numerical calculations used the resources of the Center for Scientific Computing (CSC) Frankfurt and the National Energy Research Scientific Computing Center, a DOE Office of Science User Facility supported by the Office of Science of the U.S. Department of Energy under Contract No. DE-AC02-05CH11231.

\end{document}